\documentclass[useAMS,usenatbib,usegraphicx]{mn2e}
\usepackage{amssymb}
\usepackage{latexsym}
\usepackage {keyval}
\usepackage{amsmath}
\usepackage{subfig}
\usepackage{graphicx}
\usepackage[countmax]{subfloat}
\usepackage{epsfig}
\title[H$_2$O masers in AFGL 2591 VLA~2 and VLA~3]{Multi-epoch VLBA H$_2$O maser observations toward the massive YSOs AFGL~2591 VLA~2 and VLA~3}
\author[Torrelles et al.]{J. M. Torrelles$^1$\thanks{E-mail: torrelles@ieec.cat}, M. A. Trinidad$^{2}$,
S. Curiel$^{3}$, R. Estalella$^{4}$, N. A. Patel$^5$,  
\newauthor J. F. G\'omez$^6$, G. Anglada$^6$, C. Carrasco-Gonz\'alez$^{7,8}$, J. Cant\'o$^3$, 
A. Raga$^9$, 
\newauthor L. F. Rodr\'{\i}guez$^8$
\\
$^{1}$Instituto de Ciencias del Espacio (CSIC)-UB/IEEC, Universitat de Barcelona, Mart\'{\i} i Franqu\`{e}s 1, 08028 Barcelona, Spain\\
$^{2}$Departamento de Astronom\'{\i}a, Universidad de Guanajuato, Apdo. Postal 144, 36000 Guanajuato, M\'exico\\
$^{3}$Instituto de Astronom\'{\i}a (UNAM), Apartado 70-264, 04510 M\'exico
D. F., M\'exico\\
$^{4}$Departament d'Astronomia i Meteorologia and Institut de Ci\`{e}ncies del Cosmos (IEEC-UB), Universitat de Barcelona,\\~~Mart\'{\i} i Franqu\`{e}s 1, 08028 Barcelona, Spain\\
$^{5}$Harvard-Smithsonian Center for Astrophysics, 60 Garden Street, Cambridge, MA 02138, USA\\
$^{6}$Instituto de Astrof\'{\i}sica de Andaluc\'{\i}a (CSIC), Apartado 3004, 18080 Granada, Spain\\
$^7$Max-Planck-Institut f\"ur Radioastronomie (MPIfR), Auf dem H\"ugel 69, 53121 Bonn, Germany\\
$^{8}$Centro de Radioastronom\'{\i}a y Astrof\'{\i}sica (UNAM), 58089 Morelia, M\'exico\\
$^{9}$Instituto de Ciencias Nucleares (UNAM), Apartado 70-543, 04510 M\'exico D. F., M\'exico}

\begin{document}

\date{Accepted 2013 November 7. Received 2013 October 8; in original form 2013 August 15}

\pagerange{\pageref{firstpage}--\pageref{lastpage}} \pubyear{2013}

\maketitle

\label{firstpage}

\begin{abstract}

We present multi-epoch Very Long Baseline Array (VLBA) H$_2$O maser
observations toward the massive young stellar objects (YSOs) VLA~2
and VLA~3 in the star-forming region AFGL~2591. Through these
observations, we have extended the study of the evolution of the
masers towards these objects up to a time span of $\sim$ 10~yrs,   measuring
their radial velocities and proper motions. The H$_2$O masers in VLA~3,
the most massive YSO in AFGL~2591 ($\sim$ 30--40~M$_{\odot}$),
are grouped within projected distances of $\lesssim$ 40~mas ($\lesssim$
130~AU) from VLA~3. In contrast to other H$_2$O masers in AFGL~2591,
the masers associated with VLA~3 are significantly blueshifted (up
to $\sim$ 30~km~s$^{-1}$) with respect to the velocity of the ambient
molecular cloud. We find that the H$_2$O maser cluster as a whole, 
has moved westwards of VLA~3 between the 2001 and
2009 observations, with a proper motion of $\sim$
1.2~mas~yr$^{-1}$ ($\sim$ 20~km~s$^{-1}$). We conclude that these
masers are tracing blueshifted outflowing material, shock
excited at the inner parts of a cavity seen
previously in ammonia molecular  lines and infrared images, and
proposed to be evacuated by the outflow associated with 
the massive VLA~3 source. The masers in the region
of VLA~2 are located at projected distances of $\sim$ 0.7$''$ ($\sim$
2300~AU) north from this source, with their kinematics suggesting
that they are excited by a YSO other than VLA~2. This driving
source has not yet been identified.

\end{abstract}


\section{Introduction}

AFGL~2591, located at a distance of 3.3~kpc (Rygl et al. 2012), is one of the
most extensively studied and more luminous high-mass star forming regions (e.g., van der Tak et
al. 2006; Jim\'enez-Serra et al. 2012; Sanna et al. 2012; Johnston et al. 2013; Trinidad et al. 2013). It contains
several massive young stellar objects (YSOs) over a region of $\sim$ 6$''$
($\sim$ 0.1 pc) detected at infrared and radio continuum wavelengths (VLA~1,
VLA~2, and VLA~3), but completely obscured at optical wavelengths (e.g.,
Campbell 1984; Trinidad et al. 2003). Water maser emission, which is one of the
first observed signposts of high-mass star formation, has been detected toward
VLA~2, VLA~3,  and $\sim$ 0.5$''$ ($\sim$ 1650~AU) north of VLA~3 (Tofani et
al. 1995; Trinidad et al. 2003). Very recent works by Sanna et al. (2012) and
Trinidad et al. (2013) show that the cluster of masers observed to the
north of VLA 3 (named as VLA 3-N) is formed by two bow shocks separated by
$\sim$ 0.1$''$ ($\sim$ 330~AU), and moving away from each other with velocities
of  $\sim$ 20~km~s$^{-1}$. These authors conclude that in between the two bow
shocks lie one or two still unseen YSO(s) driving the H$_2$O masers. 

VLA~3 is probably the most massive object
($\sim$ 30--40 M$_{\odot}$) dominating the infrared and mm wavelength
emission in the AFGL 2591 star-forming region (Tofani et al. 1995;
Doty et al. 2002; St\"auber et al. 2005; Jim\'enez-Serra et al.
2012; Sanna et al. 2012). VLA~3 is embedded within a high-density
($\sim$ 10$^6$ cm$^{-3}$), hot ($\sim$ 200~K) core as observed in
dust continuum and different molecular species (van der Tak 1999;
Jim\'enez-Serra et al. 2012). The molecular core shows a chemical
segregation (Jim\'enez-Serra et al. 2012), with species like H$_2$S
and $^{13}$CS peaking close ($\sim$ 0.2$''$, or $\sim$ 600 AU) to
the massive YSO, and species like  HC$_3$N, OCS, SO$_2$, SO, and
CH$_3$OH peaking at larger distances ($\sim$ 0.4$''$, or 1100 AU).
Jim\'enez-Serra et al. (2012) modelled these results through the
combination of molecular UV photodissociation and high-temperature
gas-phase chemistry within a molecular region of $\sim$ 600~AU
radius, with an innermost cavity of 120~AU radius around the central
massive object. The presence of a cavity extending westwards
of VLA~3 has also been inferred from ammonia molecular lines and
near and mid-infrared data,  showing in particular an IR-loop with
the YSO at one edge (Forrest \& Shure 1986; Torrelles et al. 1989;
Preibisch et al. 2003). It is believed that this cavity has been
evacuated by the extended east-west bipolar outflow associated with
VLA~3 and seen in CO and H$_2$, with the blueshifted outflowing
molecular gas found westwards of VLA~3 (Mitchell et al. 1992;
Tamura \& Yamashita 1992).

Very Long Baseline Interferometry (VLBI) techniques for
observing H$_2$O maser emission allow the study of the three-dimensional
velocity distribution (proper motions and radial velocity) of the
masing gas very close to the massive YSOs, with an angular resolution
better than $\sim$ 1~mas (e.g., Goddi et al. 2006; Torrelles et.
al. 2011; Chibueze et al. 2012; Kim et al. 2013). Sanna
et al. (2012)  carried out multi-epoch VLBI H$_2$O maser observations
in 2008-2009, revealing the kinematics
of the masers toward AFGL~2591 VLA~3. These observations indicate 
that the masers trace the edges of a blueshifted  expanding cavity 
created by an outflow from the central massive object.

In this work we present Very Long Baseline Array
(VLBA) H$_2$O maser observations toward the high-mass star forming
region of AFGL2591 obtained in 2001-2002 over three epochs,  
with an angular resolution of $\sim$ 0.45 mas (Section 2).  
As part of these observations, we
have already presented and discussed in a previous paper the results obtained
toward AFGL2591 VLA~3-N (Trinidad et al. 2013; hereafter Paper I).
We now concentrate on the spatio-kinematical distribution of the
cluster of masers detected toward the massive objects AFGL2591 VLA~2
and VLA~3, extending the study of the evolution of these masers
from 1999 (VLA data; Trinidad et al.  2003), 2001-2002
(VLBA data; this paper), to 2008-2009 (VLBA data; Sanna et al.
2012), covering a time span of $\sim$ 10 years of their
evolution (Section 3). The main conclusions of our studies are
presented in Section 4.

\begin{figure*}
 \centering
 \includegraphics[width=176mm, clip=true]{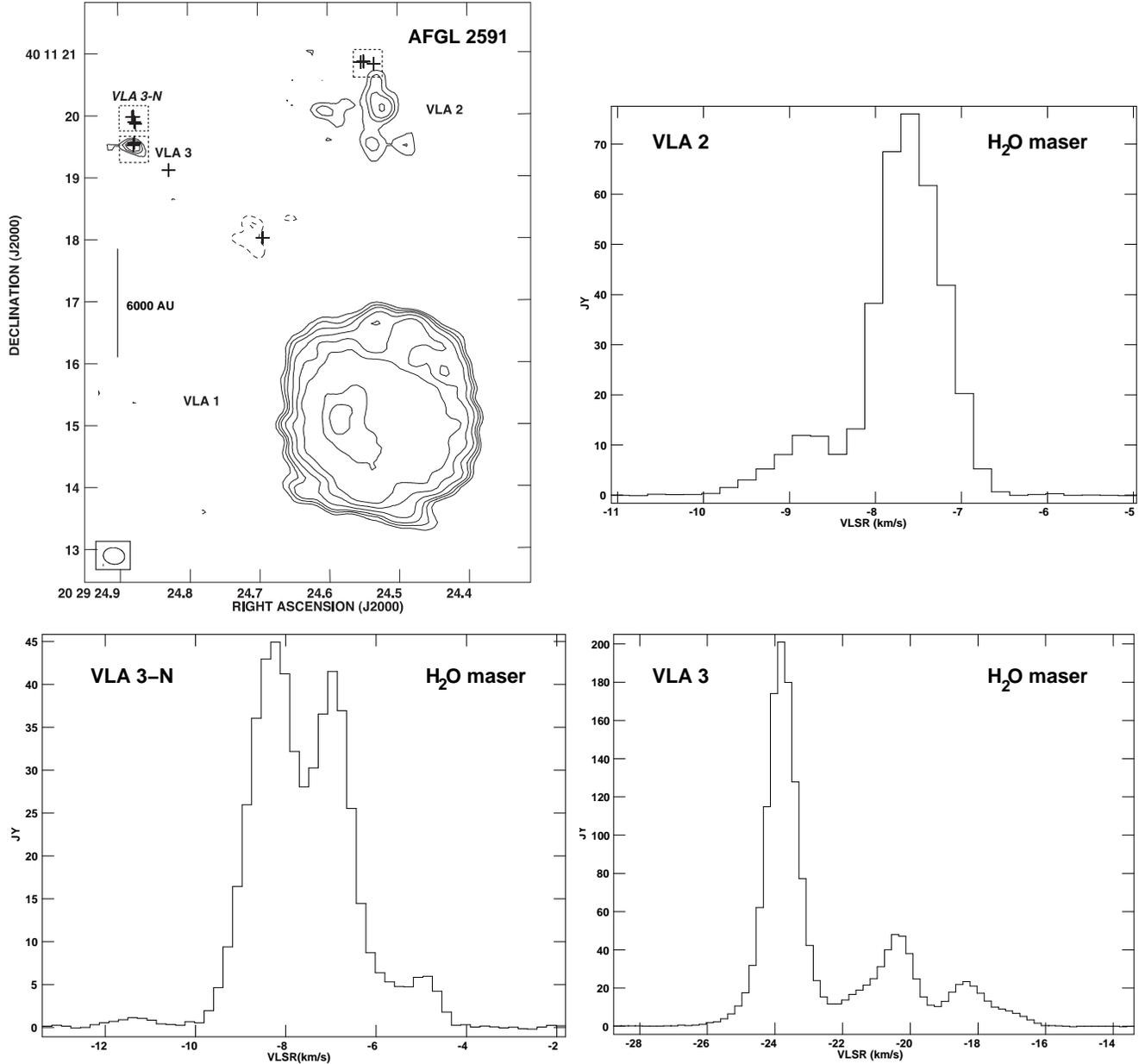} 
 \caption{VLA 3.6~cm continuum contour map of the VLA~1, VLA~2, and
 VLA~3 sources (top left panel). Plus symbols indicate the positions of the  22~GHz
 H$_2$O masers detected in the region with the VLA in epoch 1999
 June 29 (from Trinidad et al. 2003).  The H$_2$O maser spectra
 obtained with the VLBA toward the regions of VLA~2, VLA~3, and
 VLA~3-N (regions indicated in the top left panel by dashed squares)
 are also shown in the other panels for epoch 2001 December 02 (this paper). These spectra were obtained by adding components from the clean images of the regions with the task ISPEC (AIPS).
  The spatio-kinematical distribution of the H$_2$O masers as
  observed with the VLBA toward VLA~3-N is discussed by Sanna et
  al. (2012) and Paper I. These authors infer the presence there
  of a still unseen YSO(s) driving two observed H$_2$O maser bow-shock
  structures (see Section 1).
 The spatio-kinematical distribution of the masers around VLA~2 and
 VLA~3 is the subject of this present paper.}
\end{figure*}


\section{Observations}

The $6_{16}$--$5_{23}$ H$_2$O maser transition (rest frequency =
22235.08~MHz) was observed with the VLBA of the National Radio
Astronomy Observatory (NRAO)\footnote{The NRAO is a facility of the
National Science Foundation operated under  cooperative agreement
by Associated Universities, Inc.} toward AFGL~2591 at three epochs,
2001 December 2, 2002 February 11, and 2002 March 5. A bandwidth
of 8~MHz sampled over 512 channels (spectral resolution of 15.625~kHz
= 0.21~km s$^{-1}$) and centred on $V_\mathrm{LSR} = -7.6$~km~s$^{-1}$
was used, covering a radial velocity range from $V_\mathrm{LSR}
\simeq  -61$~km~s$^{-1}$  to 46~km~s$^{-1}$. 
A full description of the set-up of the VLBA observations,
amplitude and phase calibration of the observed visibilities, and further imaging of
the maser emission using the NRAO Astronomical Image Processing System (AIPS) package are extensively explained in our previous Paper I.

Several clusters of masers were detected in
the three epochs of VLBA observations with  a beam of $\sim$ 0.45~mas toward VLA~2, VLA~3, and $\sim$ 0.5$''$
north of VLA~3 (named as VLA~3-N after Paper I). As a reference, in Figure 1 we
show a 3.6 cm wavelength continuum contour map of the sources in the region
obtained with the Very Large Array (Trinidad et al. 2003). In Figure 1 we also 
show the H$_2$O maser spectra observed with the VLBA toward VLA~2, VLA~3, and
VLA~3-N in 2001 December 02 (the spectra observed with the VLBA in the two other
epochs are similar to those shown here for the first epoch). The H$_2$O maser
emission in all the AFGL~2591 region spans a radial velocity range from
$V_\mathrm{LSR} \simeq -31$ to $-2$~km~s$^{-1}$. 
More specifically, the H$_2$O maser emission is detected
in the radial velocity ranges from  $V_\mathrm{LSR} \simeq  -12$ to
$-$6~km~s$^{-1}$  in VLA~2,  $-$31 to $-16$~km~s$^{-1}$ in VLA~3,  and $-13$ to
$-2$~km~s$^{-1}$ in VLA~3-N (see Figure 1).  The rms noise level of the images
is in the range of $\sim5$~mJy~beam$^{-1}$ (channels without emission) to 
$\sim$~300~mJy~beam$^{-1}$ (channel with the strongest emission, $\sim$ 196~Jy~beam$^{-1}$ at
$V_\mathrm{LSR} \simeq -24$~km~s$^{-1}$).

We determined the position, intensity, and radial velocity of all
the maser spots in the region for the three observed epochs by
means of two-dimensional elliptical Gaussian fits. We refer to a maser
spot as emission that appears at a given velocity channel with a
signal-to-noise (S/N) ratio  $\ga$ 8 and with a distinct spatial
position for a particular epoch. The 1$\sigma$ accuracy in the
relative positions of the maser spots at each epoch is better than
$\sim$ 0.01~mas, estimated from the S/N ratio of the maser spots
and the beam size (Meehan et al. 1998). From these maser spots, we
then identified maser features in each of the observed epochs for
proper motion measurements. Here we refer  to a maser feature as a
group of maser spots coinciding simultaneously in both, position within a beam size of
$\sim$ 0.5~mas, and radial velocity within $\sim$
1~km~s$^{-1}$. 

We chose an isolated maser spot, with a point-like morphology and high intensity ($\simeq$ 10~Jy) in all our three observed epochs to self-calibrate the data and to obtain 
a first and preliminary coordinate alignment between the three epochs. 
This maser, with  $V_\mathrm{LSR}=-18.8$~km s$^{-1}$ and associated with
VLA~3, has absolute coordinates 
$\alpha$(J2000.0) = {\rm 20$^h$29$^m$24.879$^s$}, $\delta$(J2000.0)
= 40$^{\circ}$11$'$19.47$''$ ($\pm$ 0.01$''$). The procedure to align our set of VLBA epochs with the data set of Sanna et al. (2012) (epoch 2008-2009) is explained with detail in Paper I, and a brief summary is given here.  We identified a maser clearly persisting in the two sets of data (maser S17 listed by Sanna 
et al. 2012, which has a relatively small proper motion, and identified in our 
data set as maser ID28, as listed in Table 1 of Paper I). Then, we corrected the positions of the maser ID28 as a function of time (and therefore to all our data set of 2001-2002), assigning them to the expected locations of the maser S17, from a extrapolation of its position and proper motions (Sanna et al. 2012), assuming that it has moved with constant velocity through the time span of $\sim$ 7 yr. As mentioned in Paper I, the fact that after this alignment, the whole maser structure observed in epochs 2001-2002 in VLA~3-N (formed by two bow shocks separated by $\sim$ 0.1$''$) is within that reported by Sanna et al. (2012) (epochs 2008-2009), and that the estimated shift in position of our observed masers in VLA~3-N for a time span of seven years coincide with 
the angular separation between the structures observed in 2001-2002 and those of 2008-2009, gave this alignment an additional measure of robustness.

All the offset positions of the maser features shown
in the different figures of this paper (Figures 2--5) are relative
to the maser spot position (0,0) used for self-calibrating the data
of the first epoch of our VLBA observations ($\alpha$(J2000.0) = {\rm 20$^h$29$^m$24.879$^s$}, $\delta$(J2000.0)
= 40$^{\circ}$11$'$19.47$''$ ($\pm$ 0.01$''$). As mentioned above, this reference maser 
is associated with VLA 3, which has a 3.6 cm continuum
peak position 
RA (J2000) = $20^\mathrm{h}29^\mathrm{m}24.878^\mathrm{s}$, 
DEC (J2000) = $40^{\circ}11'19.49''$ 
(Trinidad et al. 2003).

In the following Section 3  we present the spatio-kinematical
distribution of the H$_2$O masers associated with VLA~2 and VLA~3.
The results on VLA~3-N were presented and discussed in Paper I.

\section{Results and discussion}

\subsection{AFGL~2591~VLA 3}

VLA~3, detected at 
infrared (Tamura et al. 1991; de Wit et al. 2009), 
mm (van der Tak et al. 1999; Jim\'enez-Serra et al. 2012), 
and
cm wavelengths (Campbell 1984; Trinidad et al. 2003; van der Tak \& Menten 2005), 
is believed to be the most massive and young object
in the AFGL 2591 star-forming region (Jim\'enez-Serra et al. 2012;
Sanna et al. 2012). This source is elongated east-west at cm continuum
wavelengths, suggesting the presence of an ionised wind driving
the east-west bipolar molecular outflow observed at large scales
($\sim$ arcminutes; Mitchell et al. 1992; Tamura \& Yamashita 1992;
Trinidad et al. 2003; van der Tak \& Menten 2005; Sanna et al.
2012; Johnston et al. 2013). The source is deeply embedded in a hot core (visual extinction
A$_V$ $\ga$ 20 mag), with molecular gas exhibiting signatures of 
Keplerian motions consistent with a mass of $\sim$ 40~M$_{\odot}$ of the
central object (van de Wiel et al. 2011; Jim\'enez-Serra et al. 2012).

We detected a cluster of masers associated with VLA~3 in our
three epochs of VLBA observations (2001 Dec 02, 2002 Feb 11, and
2002 Mar 05). The cluster is extended north-south over a scale of
$\sim$  20~mas ($\sim$ 66~AU), with a few features detected to the
north-west. In Figure 2 we show the positions of the maser features
together with their LSR radial velocity (colour scale) for our three
epochs of VLBA observations. All the velocities of the maser emission
associated with VLA~3 ($V_\mathrm{LSR} \simeq $ $-$31 to
$-16$~km~s$^{-1}$) are considerably blueshifted with respect to the
systemic velocity of the ambient molecular gas of the AFGL~2591
star-forming region ($V_\mathrm{LSR} \simeq -8$ to $-6$ km~s$^{-1}$;
van der Tak et al. 1989; Jim\'enez-Serra et al. 2012). This behaviour
differs significantly from that of the H$_2$O maser emission toward
VLA~2 and VLA~3-N, where their velocities do not have such large
blueshifted components ($V_\mathrm{LSR} \simeq  -12$ to $-$6~km~s$^{-1}$
in VLA~2, and  $-13$ to $-2$~km~s$^{-1}$ in VLA~3-N; see H$_2$O
maser spectra of these regions shown in Figure 1). The large
blueshifted components in the maser emission associated with VLA~3
were also seen in 1999 (VLA data; Trinidad et al.
2003) and 2008-2009 (VLBA data; Sanna et al. 2012), although the
maser emission detected in the 1999 and 2008-2009 epochs  is found over a somewhat
more extended region ($\sim$ 40~mas) than ours. This is shown in
Figure 2, where we also plot  the positions and radial velocities
of the H$_2$O maser features detected in 1999 (Trinidad et
al. 2003) and 2008-2009 (Sanna et al. 2012). The predominance of blueshifted (as opposed to redshifted) maser outflows has
been discussed recently by Caswell \& Phillips (2008), who conclude that the
mechanism responsible needs further exploration. Motogi et al. (2013)
propose a disc-masking scenario as the origin of the strong blue-shift dominance, 
where an optically thick disc obscures a red-shifted lobe of a compact jet. 
In the case of VLA 3, the blueshift predominance could be due to the fact that
the free-free continuum appears to be optically thick at 22 GHz (Trinidad et
al. 2003). 

The spatial alignment between our three
epochs of VLBA H$_2$O maser observations and the VLBA observations
from Sanna et al. (2012) was well determined for proper motion measurements between these two set of data, and has been extensively explained in our Paper I.
The alignment with the VLA data shown in Figure
2 (top left panel) was made  by matching the centre of the cluster
of masers observed with the VLA (epoch 1999 June 29) with the centre
of the  cluster observed with the VLBA in the closest epoch (2001
December 02). However, this geometrical centre may have suffered a
spatial displacement between these two epochs due to proper motions
of the masers. For this reason, we emphasise that the VLA maser
alignment with the VLBA data shown in all figures of this paper is only valid for
a qualitative comparison of the different data sets, but not for 
detailed studies of the proper motions.

\begin{figure*}
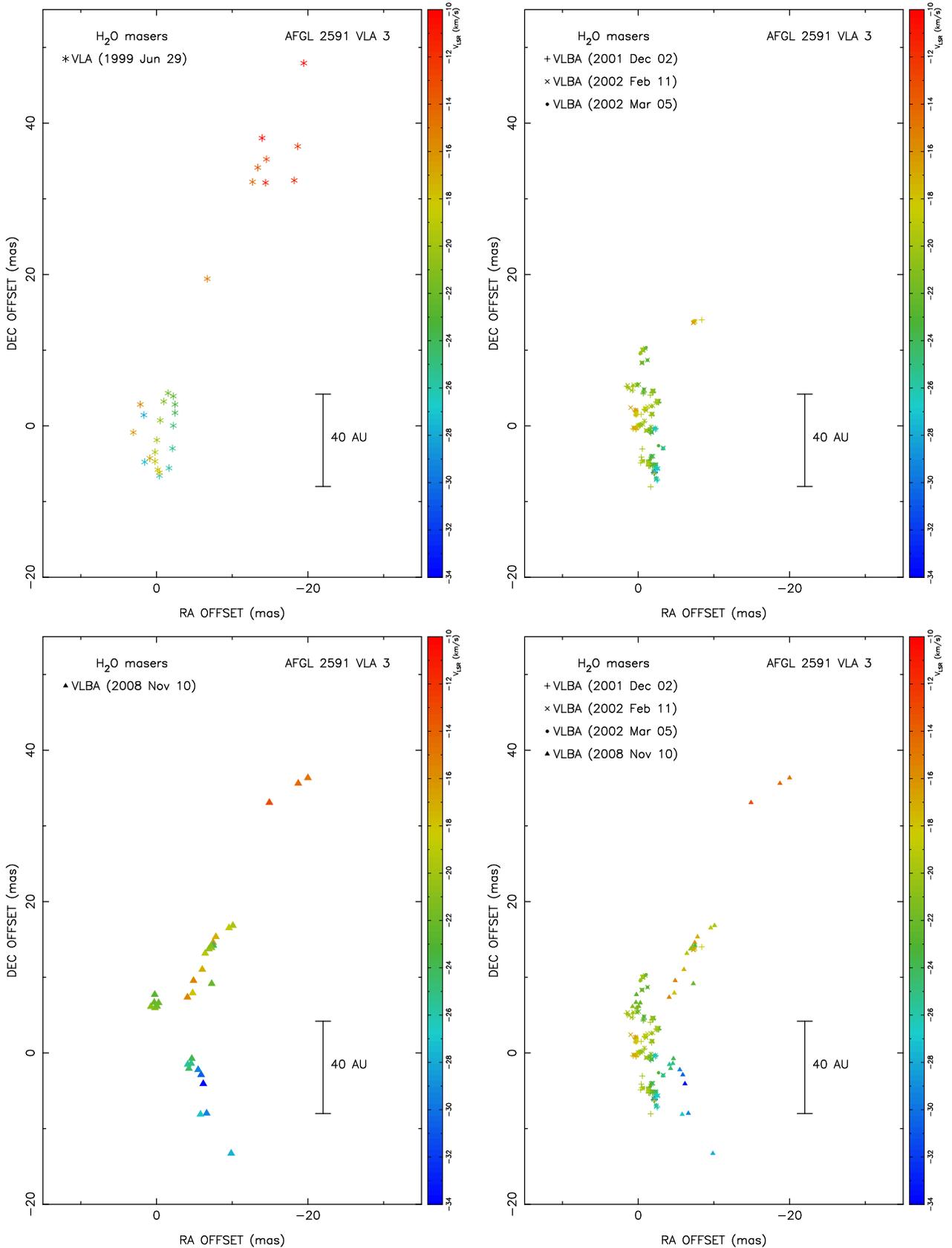


\centering
\begin{tabular}{cc}
\epsfig{file=vla3_epochv_new.eps,width=82mm,clip=true} & 
\epsfig{file=vla3_epoch123.eps,width=82mm,clip=true} \\
\epsfig{file=vla3_epocha.eps,width=82mm,clip=true} &
\epsfig{file=vla3_epoch123a.eps,width=82mm,clip=true}
\end{tabular}
\caption{Positions of the H$_2$O maser features measured with the VLA ({\it
top left:} epoch 1999 Nov 10; from Trinidad et al. 2003), VLBA ({\it
top right:} epochs 2001-2002;  this paper), and VLBA ({\it bottom
left:} epoch 2008 Nov 10; from Sanna et al. 2012) in AFGL~2591~VLA~3 (see \S~3.1).
The positions measured with the VLBA in all epochs are also shown
all together ({\it bottom right}). The colour scale represents the
radial velocity of the masers, and is the same for all the panels.  
Offset positions are relative to
the (0,0) position, 
RA(J2000) = $20^\mathrm{h}29^\mathrm{m}24.879^\mathrm{s}$, 
DEC(J2000)= $40^{\circ}11'19.47''$ 
(error $\pm0.01''$), which coincides within
the errors with the 3.6~cm continuum peak of VLA~3.} \end{figure*}

\begin{figure*}
 \centering
 \includegraphics[scale=0.90, clip=true]{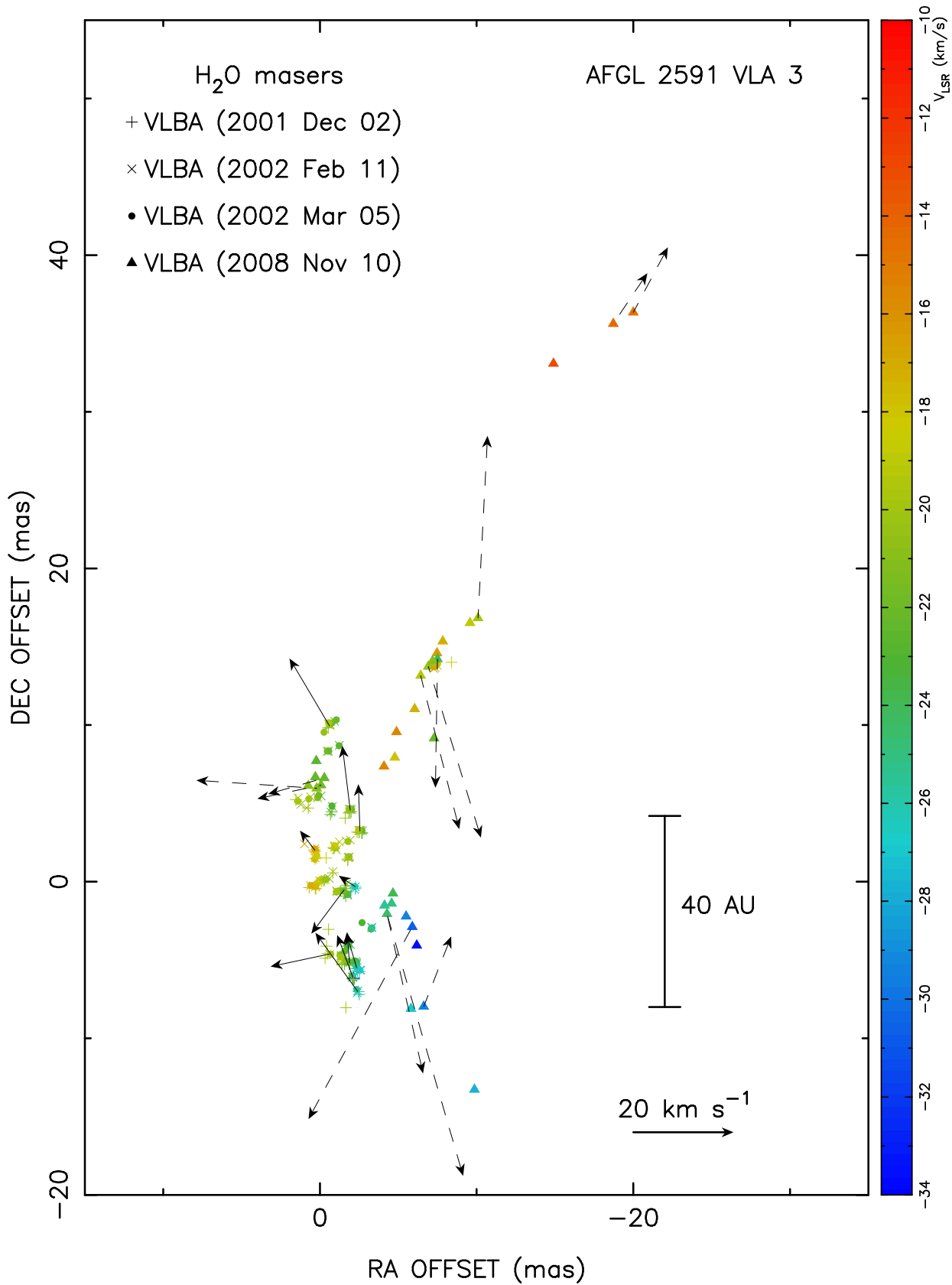} 
 \caption{Positions of the H$_2$O maser features measured with the VLBA in
AFGL~2591~VLA~3 for the  epochs 2001 Dec 02 (plus signs), 2002 Feb
11 (cross signs), and 2002 Mar 05 (filled circle signs).  Solid
arrows represent the proper motion vectors of the maser features
measured with the VLBA in these three epochs.  The position of the
H$_2$O masers (filled triangles) and proper motions (dashed arrows)
of the features measured with the VLBA by Sanna et al. (2012) in
epochs 2008-2009 are also shown. The colour scale represents the radial
velocity of the masers, and is the same as shown in Figure 2. The scale of the proper motion vector
magnitude is indicated.} \end{figure*}

\begin{figure*}
 \centering
 \includegraphics[scale=0.7]{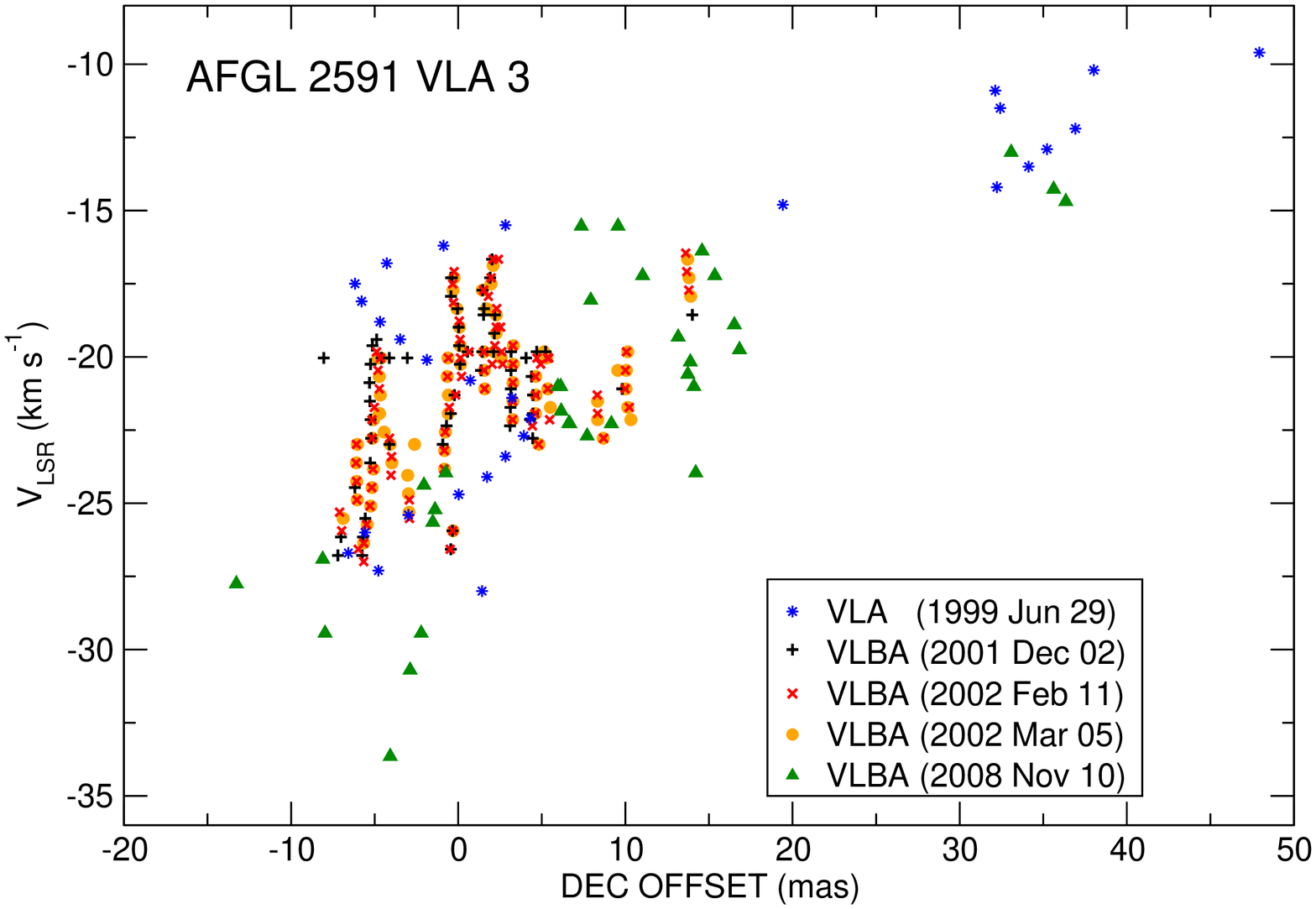} 
 \caption{Radial velocity  as a function of declination offset of
 the H$_2$O maser features measured with the VLA and VLBA from 1999 to 2008,
 showing a north-south velocity gradient (see \S\S~2 and 3.1 for the alignment of the different epochs). All the maser emission is
 significantly blueshifted with respect to the velocity of the ambient molecular
 gas, $V_\mathrm{LSR} \simeq -6$ to $-8$ km~s$^{-1}$ (van der Tak et al. 1989;
 Jim\'enez-Serra et al. 2012).}

\end{figure*}

From Figure 2 we see that there is a displacement of the H$_2$O
maser structure as a whole toward the west between the 2001-2002
and 2008-2009 epochs. The observed displacement of $\sim$ 9~mas
corresponds to a proper motion of $\sim$ 1.2~mas~yr$^{-1}$ ($\sim$
20~km~s$^{-1}$) in a time span of seven years. On the other hand,
from about 80 H$_2$O maser features that we have found in each of
our three observed epochs (2001-2002), we have only been able
to identify 11 maser features that persist in the three epochs to
measure their individual proper motions. With persisting masers we
refer to those having similar positions (within a few mas) and radial
velocities (within 1$-$2~km~s$^{-1}$) in the three different observed
epochs. The resulting proper motions of these 11 persisting
maser features are shown in Figure 3 with solid arrows. In 
Figure 3 we also show the proper motion vectors of 13 H$_2$O maser
features measured by Sanna et al. (2012) for epochs 2008-2009 (dashed
arrows). We have not been able to identify any individual maser feature persisting from epochs 2001-2002 to 2008-2009 epochs (in contrast to what we find in VLA 3-N, with individual maser features persisting over seven years; Paper I).
The magnitudes of the individual proper motions of the
maser features range from $\sim$ 0.5 to 3.6~mas~yr$^{-1}$ ($\sim$
8 to 60~km~s$^{-1}$). However, we find that these vectors do not
show a preferential motion direction in the sky. This indicates
that while the whole structure of masers is moving westwards,
within it there are individual H$_2$O masers moving in
a chaotic way, with a very short lifetime in comparison to the scale
of seven years, when the motion of the full structure becomes apparent.
This would be consistent with the fact that only a small percentage
of the maser features identified in the region have been found to
persist in our three consecutive epochs of observations, implying
lifetimes for most of them of $\lesssim$ 3~months (see also below).

Sanna et al. (2012) identified in their VLBA data (epochs 2008-2009) a V-shaped
structure of the H$_2$O maser distribution opening toward the west, with its
vertex close to the peak of the radio continuum emission of VLA~3 (position
$[0,0]$ in Figure 2; bottom left panel).  This structure is not observed
in our VLBA observations (2001-2002), where the cluster of masers is
concentrated closer to VLA~3 (see Figure 2; top right panel). 
However, we think that the V-shaped structure is
not a new structure that has appeared after  a
time span of seven years. In fact, the VLA H$_2$O maser observations
by Trinidad et al. (2003) also shows signs that it
was already there in 1999, as indicated by the masers
detected then with positions $\sim$ ($-15$~mas, +35~mas) and radial
velocities $V_\mathrm{LSR} \simeq -$11~km~s$^{-1}$ (see Figure 2;
top left panel). These masers have similar positions and velocities
than the VLBA masers of epochs 2008-2009  that constitute the northern
part of the V-shaped structure (see Figure 2; bottom left panel). We therefore
think that the absence of the V-shaped structure in the
2001-2002 epochs is just due to the well known high time
variability of the flux density of the H$_2$O masers.

An additional, important result is that the masers in all the
observed epochs present a radial velocity gradient from north
to south, from $V_\mathrm{LSR} \simeq -$12~km~s$^{-1}$ (northern
regions) to $-$30~km~s$^{-1}$  (southern regions), as shown in
Figure 4. This velocity gradient ($\sim$ 0.12~km~s$^{-1}$~AU$^{-1}$) was also noticed by Trinidad et
al. (2003) and Sanna et al. (2012) in their respective epochs of
observations, although by combining all the masers detected in all
the epochs as shown in Figure 4, this gradient is seen more clearly. However, we notice that the VLBA data do not show the peculiar velocity-position helical distribution seen with the VLA in 1999 (Trinidad et al. 2003).

\begin{figure*}
 \centering
 \includegraphics[width=176mm, clip=true]{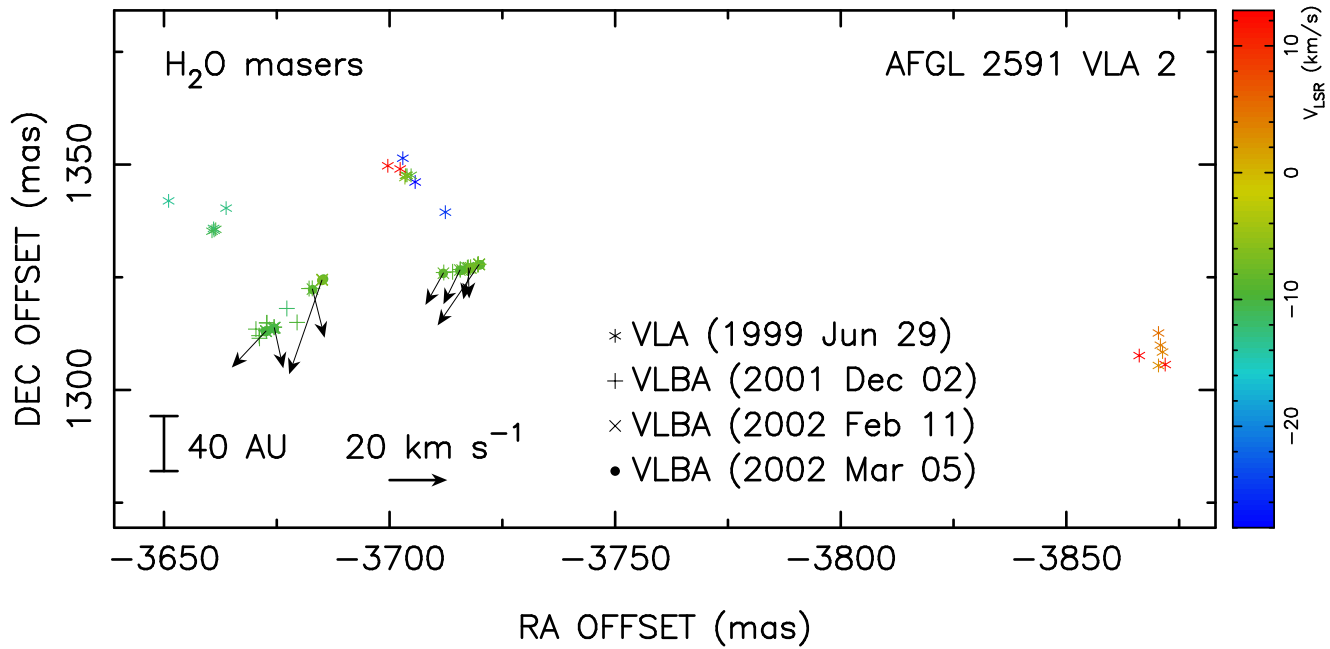} 
 \caption{Positions of the H$_2$O maser features measured with the VLBA
 toward
AFGL~2591~VLA~2 (see Fig. 1) for the  epochs 2001 Dec 02 (plus
signs), 2002 Feb 11 (cross signs), and 2002 Mar 05 (filled circles).  Solid arrows represent the proper motion vectors of the
maser features measured with the VLBA in these three epochs.  The
position of the H$_2$O maser features measured with the VLA by
Trinidad et al. (2003) in epoch 1999 Jun 29 are also shown (asterisks; see \S\S ~2 and 3.1 for the alignment of the different epochs). The colour scale represents the radial velocity of the masers.
The scale of the proper motion vector magnitude is indicated.  VLA~2
is at the offset position $\sim$ ($-$3945~mas, $+$672~mas), this
is $\sim$ 0.7$''$ ($\sim$ 2310~AU) south from the detected masers shown here.}
 \end{figure*}

The large values of the velocities that we find in the H$_2$O masing
region of VLA~3 with respect to the ambient cloud velocity, with total velocities (tangential + radial velocities)
$\gtrsim$ 40~km~s$^{-1}$, indicate that the masers are tracing
outflowing material. In fact, the central mass required  to gravitationally
bind these motions would exceed 200~M$_{\odot}$,
which is not observed. The westward proper motion of the
whole H$_2$O maser structure associated with VLA~3, in a time span of seven years (Figure 2), together with their
large blueshifted velocities with respect to the ambient  local medium,
suggest a physical relationship between the masers and the blueshifted
motions of the large-scale outflow seen in CO and H$_2$, found also
toward the west (Mitchell et al. 1992; Tamura \& Yamashita 1992).
Furthermore, the H$_2$O masers are spatially distributed within a
region of $\sim$ 40~mas around the radio continuum emission peak of VLA~3
(Figures 1 and 2), which is located at the edge of the cavity
observed through molecular lines and infrared images extending
$\sim$ 10$''$ ($\sim$ 3.3$\times$10$^3$~AU) toward the west (Forrest \& Shure 1986; Torrelles et
al. 1989; Preibisch et al. 2003). Due to all this, we propose that
the H$_2$O masers are tracing the inner parts of  the molecular
walls of that cavity, very close ($\lesssim$ 130 AU) to the massive
YSO, and shock excited and evacuated by the east-west outflow,
marking a strong interaction between the outflowing gas from VLA3 and
the surrounding molecular gas (the H$_2$O maser emission would occur behind shocks propagating in dense regions; Hollenbach, Elitzur, \& McKee 2013). This strong interaction could also
contribute to the high temperatures observed toward VLA~3 ($\gtrsim$
200~K) and measured through  ammonia emission (Torrelles et al.
1989). 

A rough estimate of upper limits for the coherent path of the maser emission could be derived if we assume unsaturated tubular masers with beaming angles  $\Delta$$\Omega$ $= (d/l)^2$, where $d$ and $l$ are the maser transverse and longitudinal sizes, respectively (e.g., Vlemmings et al. 2006; Surcis et al. 2011; Hollenbach et al. 2013). As pointed out by Hollenbach et al. (2013), beaming angles themselves are unmeasurable and can only be inferred indirectly using different approaches. In this sense, Vlemmings et al. (2006) and Surcis et al. (2011), through polarisation measurements of the H$_2$O maser emission and using full radiative transfer codes (based on the models for H$_2$O masers of Nedoluha \& Watson 1992), obtain beaming angles 10$^{-2}$ $\lesssim$ $\Delta$$\Omega$ $\lesssim$ 0.5 in different high-mass star-forming regions. Assuming these beaming angles, and a transverse size $d$ $\lesssim$ 0.45~mas ($\lesssim$ 1.5~AU, unresolved masers) we obtain a path length $l$ $\lesssim$ 15~AU. Considering that the size of the cavity in VLA~3 is $\sim$ 3.3$\times$10$^3$~AU, we think that the cavity walls could have enough thickness to provide such a path of coherent velocities (see also Uscanga et al. 2005).

As mentioned above, all the data available and presented in this
work reveal a radial velocity gradient in the H$_2$O masers along
the north-south direction, with higher blueshifted velocities to
the south with respect to the velocity of the ambient molecular gas
(Figure 4). Interestingly, Tamura \& Yamashita (1992) also
find an extinction gradient along the north-south direction of the
IR-loop delineating the cavity  of VLA~3 (Forrest \& Shure 1986),
with a higher extinction in the north part of the IR-loop. We think
that this north-south asymmetry, with a lower amount of gas to the
south relative to the north, could produce higher expansion velocities
to the south due to the interaction of outflowing material from
VLA~3 with this non-uniform ambient medium, explaining in this way
the observed radial velocity gradient in the masers.  However, we 
cannot rule out other effects that could produce
the observed north-south  velocity gradient in the H$_2$O masers
(e.g., shock-excited ambient molecular gas accelerated by a wind
from a precessing source, as suggested by Trinidad et al. 2003).

\subsection{AFGL~2591~VLA 2}

We detected H$_2$O maser emission toward VLA~2 in our three
epochs of VLBA observations. The VLBA H$_2$O maser observations
reported by Sanna et al. (2012) did not cover the region of VLA~2,
so unfortunately in this case we are not able to extend the study
of the evolution of the masers beyond our epochs of 2001-2002, as
it was done in the case of VLA~3 (Section 3.1).

In Figure 5 we show the positions of the H$_2$O maser features
together with their proper motions as measured with our VLBA
observations. The detected H$_2$O masers are located $\sim$ 0.7$''$
($\sim$ 2300~AU) north from the peak position of the 3.6~cm continuum
emission of VLA~2. The proper motions of the H$_2$O masers are
toward the south, with values of $\sim$ 20--40~km~s$^{-1}$.
We identify the clusters of masers detected with the VLBA with those
found with the VLA in 1999 (see Figures 5 and 1).

The radio continuum emission of VLA~2 is consistent with a partially
optically thick HII region excited by an early B-type star (Trinidad
et al. 2003). The lack of any symmetry in the spatio-kinematical
distribution of the H$_2$O masers detected in this region, together
with the lack of additional physical information on the characteristics
of VLA~2 (e.g., there is no apparent outflow activity associated
with this source), makes difficult to establish a firm conclusion
on a possible H$_2$O maser-VLA~2 physical association. A possibility
would be that the H$_2$O masers moving southwards are tracing
infalling gas motions around VLA~2, but this would require a central mass of $\gtrsim$
10$^{3}$~M$_{\odot}$ to bind motions of $\gtrsim$~20~km~s$^{-1}$
at distances of $\sim$ 2300~AU.  This is the reason why we rather
favour that the observed H$_2$O masers in this region are excited
by a nearby YSO that has yet to be identified. 
We think that new high-sensitive cm
and (sub)mm wavelength continuum observations in the region could
clarify this issue.

\section{Conclusions}

We present and analyse multi-epoch VLBA H$_2$O maser observations
toward the massive YSOs AFGL~2591 VLA~2 and VLA~3. By comparing our
data with those previously reported by Trinidad et al. (2003) and
Sanna et al. (2012), we have extended the study of the kinematics
of the H$_2$O masers in these two regions up to a time span of
$\sim$ 10~yrs.  The cluster of masers found within $\sim$ 40~mas
($\sim$ 130~AU) from the most massive object in this star-forming
region, VLA~3 ($\sim$ 30--40~M$_{\odot}$), is significantly blueshifted
with respect to the ambient molecular cloud, and moving as a whole
toward the west of VLA~3 with a proper motion of $\sim$
20~km~s$^{-1}$. VLA~3 and its associated  H$_2$O masers are located
at the edge of a cavity of $\sim$ 10$''$ size extending toward the
west and seen previously through ammonia lines and infrared images.
We propose that the masers are tracing blueshifted outflowing
material, shock excited at the inner parts of that cavity
evacuated by the outflow of VLA~3. This interpretation is fully
consistent with the one proposed previously by Sanna and collaborators.
In addition, we find a radial velocity gradient in the H$_2$O masers
along the north-south direction, with higher blueshifted velocities
to the south. We interpret this radial velocity gradient as due to
the decrease of the amount of gas to the south, needed to explain
the observed lower extinction in the southern parts. 

On the other
hand, the spatio-kinematical distribution of the H$_2$O masers in
the region of VLA~2 favours that these masers are excited by a YSO
other than VLA~2. Future, sensitive radio continuum observations
could help to discover this new driving source, as well as the exciting source of the VLA~3-N system (Sanna
et al 2012; Trinidad et al. 2013).

\section*{Acknowledgments}

We would like to thank our referee for the very useful report on our manuscript. 
We also acknowledge Wouter Vlemmings and Moshe Elitzur for very helpful comments.
GA, CC-G, RE, JFG, and JMT acknowledge support from MICINN (Spain)
AYA2008-06189-C03 and AYA2011-30228-C03 grants (co-funded with FEDER funds). JC and ACR
acknowledge support from CONACyT grant 61547.  SC acknowledges
the support of DGAPA, UNAM, CONACyT (M\'exico) and CSIC (Spain).
LFR acknowledges the
support of DGAPA, UNAM, and of CONACyT (M\'exico).  MAT acknowledges
support from CONACyT grant 82543. RE and JMT acknowledge support
from AGAUR (Catalonia) 2009SGR1172 grant. JMT acknowledges the hospitality offered by the Science Operations Center of the NRAO (Socorro, NM; USA) where part of the data reduction was carried out (June-July, 2013). The ICC (UB) is a
CSIC-Associated Unit through the ICE (CSIC).


\label{lastpage}

\bsp


\begin{thebibliography}{99}

\bibitem[\protect\citeauthoryear{Campbell}{1984}]{cam84} Campbell B. 1984, ApJ, 287, 334
\bibitem[\protect\citeauthoryear{Caswell 
\& Phillips}{2008}]{cas08} Caswell J.~L., Phillips C.~J.\
2008, MNRAS, 386, 1521
\bibitem[\protect\citeauthoryear{Chibueze et al.}{2012}]{chi12} Chibueze J. O., Imai H., Tafoya D., Omodaka T., Kameya O., Hirota T., Chong S., Torrelles J. M. 2012, ApJ, 748, 146
\bibitem[\protect\citeauthoryear{de Wit et al.}{2009}]{dew09} de Wit W. J., Hoare M. G., Fujiyoshi T. et al. 2009, A\&A, 494, 157
\bibitem[\protect\citeauthoryear{Doty et al.}{2002}]{dot02} Doty S. D., van Dishoeck E. F., van der Tak F. F. S., Boonman A. M. S. 2002, A\&A, 389, 446
\bibitem[\protect\citeauthoryear{Forrest \& Shure}{1986}]{for86} Forrest W. J., Shure M. A. 1986, ApJ, 311, L81
\bibitem[\protect\citeauthoryear{Goddi et al.}{2006}]{god06} Goddi C., Moscadelli L., Torrelles J. M., Uscanga L., Cesaroni R. 2006, A\&A, 447, L9
\bibitem[\protect\citeauthoryear{Hollenbach et al.}{2013}]{hol13} Hollenbach D., Elitzur M., McKee C. F. 2013, ApJ, 773, 70
\bibitem[\protect\citeauthoryear{Jim\'enez-Serra et al.}{2007}]{jim07} Jim\'enez-Serra I., Mart\'{\i}n-Pintado J., Rodr\'{\i}guez-Franco A., Chandler C., Comito C., Schilke P. 2007, ApJ, 661, L187
\bibitem[\protect\citeauthoryear{Jim\'enez-Serra et al.}{2012}]{jim12} Jim\'enez-Serra I., Zhang Q., Viti S., Mart\'{\i}n-Pintado J., de Wit W.-J. 2012, ApJ, 753, 34
\bibitem[\protect\citeauthoryear{Johnston et al.}{2013}]{jo13} Johnston K. G., Shepherd D. S., Robitaille T. P., Wood K. 2013, A\&A, 551, A43 
\bibitem[\protect\citeauthoryear{Kim et al.}{2013}]{kim13} Kim J-S., Kim S-W., Kurayama T., Honma M., Sasao T., Surcis G., Cant\'o J., Torrelles J. M., Kim S. J. 2013, ApJ, 767, 86
\bibitem[\protect\citeauthoryear{Meehan et al.}{1998}]{mee98} Meehan L. S. G., Wilking B. A., Claussen M. J., Mundy L. G., Wootten A. 1998, AJ, 115, 1599
\bibitem[\protect\citeauthoryear{Mitchell et al.}{1992}]{mit92} Mitchell G. F., Hasegawa T. I., Schella J. 1992, ApJ, 386, 604
\bibitem[\protect\citeauthoryear{Motogi et al.}{2013}]{mot13} Motogi K., Sorai K., Niinuma K., et al.\ 2013, MNRAS, 428, 349 
\bibitem[\protect\citeauthoryear{Nedoluha \& Watson}{1992}]{ned92} Nedoluha G. E., Watson W. D. 1992, ApJ, 384, 185
\bibitem[\protect\citeauthoryear{Preibisch et al.}{2003}]{pre03} Preibisch T., Balega Y. Y., Schertl D., Weigelt G. 2003, A\&A, 412, 735
\bibitem[\protect\citeauthoryear{Rygl et al.}{2012}]{ryg12} Rygl K. L. J., Brunthaler A., Sanna A., Menten  K. M.,  Reid M. J., van Langevelde H. J., Honma M., Torstensson K. J. E., Fujisawa K. 2012, A\&A, 539, 79
\bibitem[\protect\citeauthoryear{Sanna et al.}{2012}]{san12} Sanna A., Reid M. J., Carrasco-Gonz\'alez C., Menten K. M., Brunthaler A., Moscadelli L., Rygl K. L. J. 2012, ApJ, 745, 191
\bibitem[\protect\citeauthoryear{St\"auber et al.}{2005}]{sta05} St\"auber P., Doty S. D., van Dishoeck E. F., Benz A. O. 2005, A\&A, 440, 949
\bibitem[\protect\citeauthoryear{Surcis et al.}{2011}]{sur11} Surcis G., Vlemmings W. H. T., Torres, R. M., van Langevelde H. J., Hutawarakorn Kramer, B. 2011, A\&A, 533, A47
\bibitem[\protect\citeauthoryear{Tamura et al.}{1991}]{tam91} Tamura M., Gatley I., Joyce R. R., Ueno M., Suto H., Sekiguchi M. 1991, ApJ, 378, 611 
\bibitem[\protect\citeauthoryear{Tamura \& Yamashita}{1992}]{tam92} Tamura M., Yamashita T. 1992, ApJ, 391, 710
\bibitem[\protect\citeauthoryear{Tofani et al.}{1995}]{tof95} Tofani G., Felli M., Taylor G. B., Hunter T. R. 1995, A\&AS, 112, 299
\bibitem[\protect\citeauthoryear{Torrelles et al.}{1989}]{tor89} Torrelles J. M., Ho P. T. P., Rodr\'{\i}guez L. F., Cant\'o J. 1989, ApJ, 343, 222
\bibitem[\protect\citeauthoryear{Torrelles et al.}{2011}]{tor11} Torrelles J. M., Patel N. A., Curiel S., 
     Estalella R., G\'omez J. F., Rodr\'{\i}guez L. F., Cant\'o J., Anglada G.,
     Vlemmings W., Garay G., Raga A. C., Ho P. T. P. 2011, MNRAS, 410, 627
\bibitem[\protect\citeauthoryear{Trinidad et al.}{2003}]{tri03} Trinidad M. A., Curiel S., Cant\'o J., D'Alessio P., Rodr\'{\i}guez L. F., Torrelles J. M., G\'omez J. F., Patel N., Ho P. T. P.
     2003, ApJ, 589, 386
\bibitem[\protect\citeauthoryear{Trinidad et al.}{2013}]{tri13} Trinidad M. A., Curiel S., Estalella R., Cant\'o J., Raga A., Torrelles J. M., Patel N. A., G\'omez J. F., Anglada G., Carrasco-Gonz\'alez C., Rodr\'{\i}guez L. F. 2013, MNRAS, 430, 1309 [Paper I]
\bibitem[\protect\citeauthoryear{Uscanga et al.}{2005}]{usc05} Uscanga L., Cant\'o J., Curiel S., Anglada G., Torrelles J. M., Patel N. A., G\'omez J. F., Raga A. C. 2005, ApJ, 634, 468
\bibitem[\protect\citeauthoryear{van der Tak \& Menten}{2005}]{van05} van der Tak F. F. S., Menten K. 2005, A\&A, 437, 947
\bibitem[\protect\citeauthoryear{van der Tak et al.}{1999}]{van99} van der Tak F. F. S., van Dishoeck
     E. F., Evans II N. J., Bakker E. J., Blake G. A. 1999, ApJ, 522, 991
\bibitem[\protect\citeauthoryear{van der Tak et al.}{2006}]{van06} van der Tak F. F. S., Walmsley C. M., Herpin F., Ceccarelli, C. 2006, A\&A, 447, 1011
\bibitem[\protect\citeauthoryear{van der Wiel et al.}{2011}]{van11} van der Wiel M. H. D., van der Tak F. F. S., Spaans M., Fuller G. A., Plume R., Roberts H., Williams J. L. 2011, A\&A, 532, 88
\bibitem[\protect\citeauthoryear{Vlemmings et al.}{2006}]{vle06} Vlemmings W. H. T., Diamond, P. J., van Langevelde H. J., Torrelles J. M. 2006, A\&A, 448, 597


\end{thebibliography}
\end{document}